\documentclass[11pt,dvips]{article}
\usepackage{epsfig}
\usepackage{times}
\makeatletter
\renewcommand{\section}{\@startsection{section}{1}{0in}
	{0.4\baselineskip}{0.1\baselineskip}{\Large\bf}}
\renewcommand{\subsection}{\@startsection{subsection}{2}{0in}
	{0.25\baselineskip}{-\baselineskip}{\large\bf}}
\renewcommand{\subsubsection}{\@startsection{subsubsection}{3}{0in}
	{0.1\baselineskip}{-\baselineskip}{\normalsize\bf}}
\makeatother
\begin{document}

%  Session and Paper Code:
%\thispagestyle{myheadings}
%
%  ***INSTRUCTIONS:***  Replace `OG 9.9.9' in the command argument below
%			with your assigned session and paper code:
%\markright{HE 4.1.07}
%
\makeatletter\newcommand{\ps@icrc}{
\renewcommand{\@oddhead}{\slshape{HE.4.1.07}\hfil}}
\makeatother\thispagestyle{icrc}
\begin{center}
{\LARGE \bf Neutrino Oscillations at High Energy by MACRO }
\end{center}

%  Author List:
\begin{center}
{\bf F Ronga$^{1}$ for the MACRO collaboration}
\\
{\it $^{1}$INFN Laboratori Nazionali di Frascati P.O. Box 13 Frascati ITALY\\}
\end{center}

%  Abstract:
\begin{center}
{\large \bf Abstract\\}
\end{center}
\vspace{-0.5ex}
%
%  ***INSTRUCTIONS:***  Replace text below with your own abstract:
%
We present updated results of the measurement of upward-going muons produced by neutrino interactions
in the rock below the MACRO detector. These data support MACRO's previously
published results. 
They favor a neutrino oscillation explanation of the atmospheric neutrino anomaly.

\section{Introduction:}
\label{intro.sec}
 The interest in precise measurements of the flux of neutrinos produced
in cosmic ray cascades in the atmosphere has been growing over the last years
due to the anomaly in the ratio of contained muon neutrino to electron neutrino
 interactions.
The past observations of Kamiokande, IMB and Soudan 2  are now  confirmed by
those of SuperKamiokande, MACRO and Soudan2 (with higher statistics) 
and the anomaly finds explanation in
the scenario of $\nu_{\mu}$ oscillation (Fukuda 1998a).
The effects of neutrino oscillations have to appear also in higher energy ranges. The flux of
muon neutrinos in the energy region from a few GeV up to a few TeV can be inferred from
measurements of upward throughgoing muons (Ahlen 1995,Ambrosio 1998b,Hatakeyama 1998,
Fukuda 1998b). As a consequence of oscillations, the flux of upward throughgoing muons should be
affected both in the absolute number of events and in the shape of the zenith angle distribution, with
relatively fewer observed events near the vertical than near the horizontal due to the longer path
length of neutrinos from production to observation.

Here an update of the measurement of the high energy muon neutrino flux is presented. The new data
are in agreement with the old data. The MACRO low energy data are presented in another paper at this
conference (Surdo 1999).

\begin{figure} [t] 
\epsfig{file=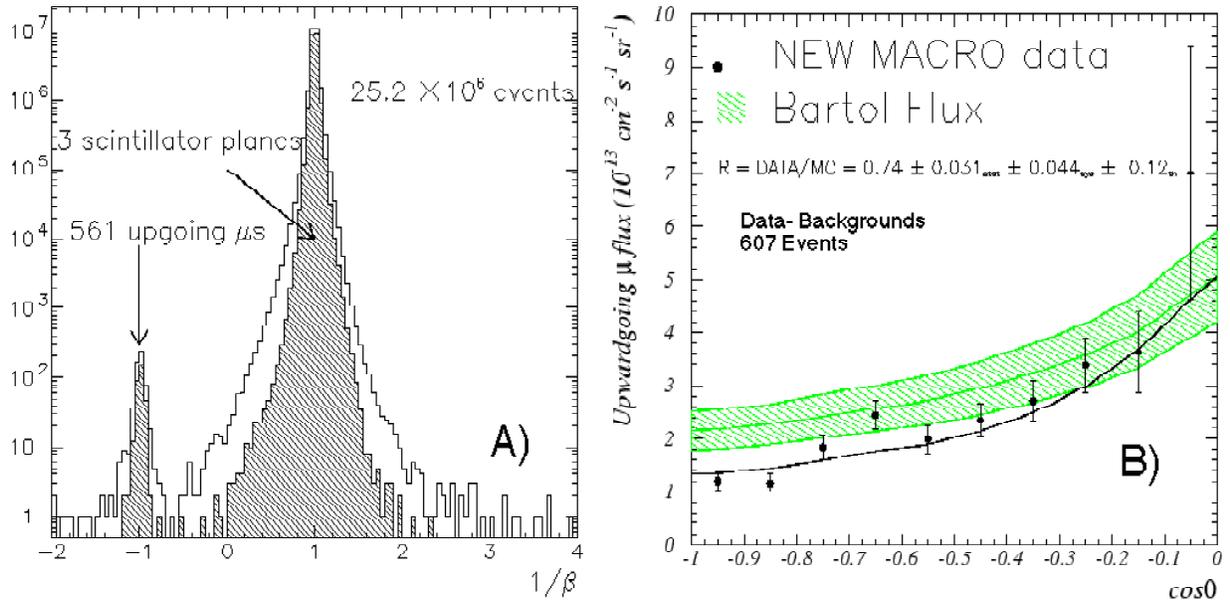,height=80mm}
%\psfig{file=fig1.eps,height=80mm}
%\indent{\picture 110mm by 80mm (Fig1 scaled 450)}
\caption {\label{fig:topo}  A)  Distribution of $1/\beta$ for the full detector data set. A clear
peak of upward muons is evident centered at $1/\beta =-1$. The widths of the
distributions for upgoing and downgoing muons are consistent. The
shaded part of the distribution is for the subset of events where
three scintillator layers were hit. B) Zenith distribution of flux of upward throughgoing
muons with energy greater than 1 GeV for data and Monte Carlo for the combined MACRO
data. The shaded region shows the
expectation for no oscillations and includes 
the 17\% uncertainty in the expectation.
The lower line shows the prediction for an oscillated flux with
$\sin^2 2 \theta = 1$ and $\Delta m^2 = 0.0025$ eV$^2$}
\end{figure}

\section{Upward Throughgoing Muons:}
The MACRO detector is described elsewhere (Ahlen 1993, Ambrosio 1998b).
Active elements are streamer tube chambers used for tracking and liquid scintillator counters
used for the time measurement.
The direction that muons travel through MACRO is determined by the
time-of-flight between two different layers of scintillator counters.
 The measured muon velocity is calculated with the
convention that muons going down through the detector are expected to have
1/$\beta$ near +1 while muons going up through the detector are expected
to have 1/$\beta$ near -1.

Several cuts are imposed to remove backgrounds caused by radioactivity or showering events
which may result in bad
time reconstruction. The most important cut requires that the
position of a muon hit in each scintillator as determined from the
timing within the scintillator counter agrees 
within $\pm$70 cm with the position
indicated by the streamer tube track.

When a muon hits 3 scintillator layers, there is  redundancy in the
time measurement and 1/$\beta$ is calculated from a linear fit of the times
as a function of the pathlength. Tracks with a
poor fit are rejected. Other minor cuts are applied for 
 the tracks with only two layers of scintillator hit.

It has been observed that downgoing muons which pass near or through
MACRO may produce low-energy, upgoing particles.  These could appear to be
neutrino-induced
upward throughgoing muons if the down-going muon misses the detector (Ambrosio 1998a).
In order to reduce this background, we impose a cut
requiring that each upgoing muon must cross at least 200 g/cm$^2$ of
material in the bottom half of the detector. Finally, a large number
of nearly horizontal ($\cos \theta > -0.1$), but upgoing muons have
been observed coming from azimuth angles corresponding to a  direction containing 
a cliff in the mountain where the overburden is insufficient to remove
nearly horizontal, downgoing muons which have scattered in the mountain
and appear as upgoing. We exclude this region from both our
observation and Monte-Carlo calculation of the upgoing events.
 
\begin{figure} [ht]
\epsfig{file=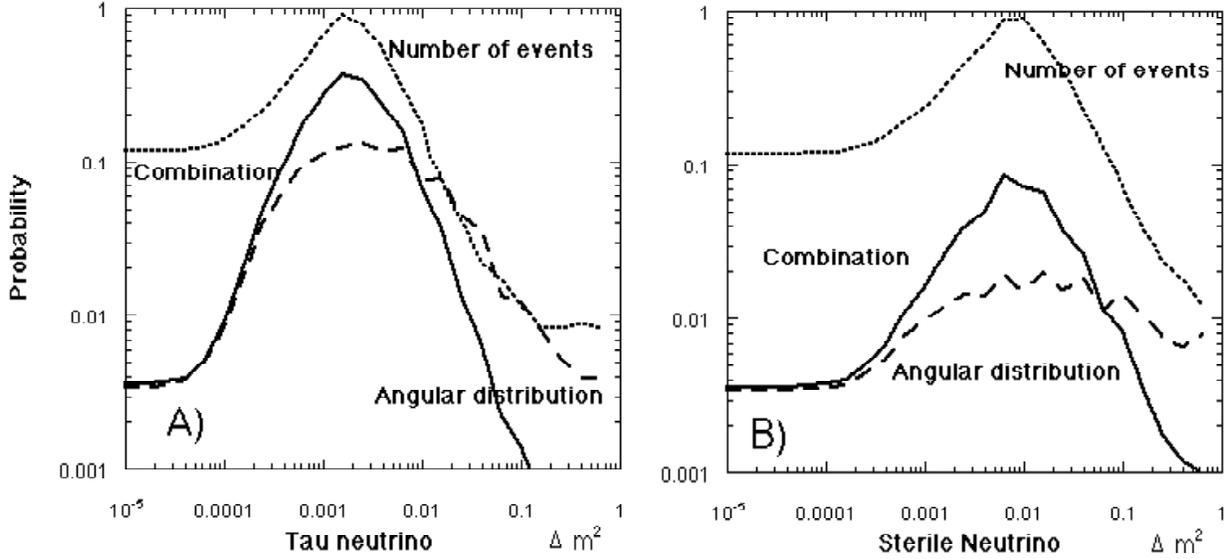,height=80mm}
%\psfig{file=fig3.eps,height=75mm}
%\indent{\picture 110mm by 75mm (Fig3 scaled 600)}
\caption{  \label{fig:prob}  Probabilities for maximum mixing and  oscillations 
 $\nu_{\mu} \rightarrow\nu_{\tau}$ (Fig 2 A)
or oscillations $\nu_{\mu} \rightarrow$ sterile neutrino (Fig 2 B). 
The 3 lines corresponds to the probability
from the total number of events (dotted line), the probability from the chi-square of 
the angular distribution with data
and prediction normalized (dashed line)  and to the combination of the two 
independent probabilities (continous line). }
\end{figure}
 
  Figure 1A) shows the
$1/\beta$ distribution for the throughgoing data from the full detector running. 
A clear peak of upgoing muons is evident centered on $1/\beta=-1$.

There are 561 events in the range $-1.25 < 1/\beta < -0.75$ which we
define as upgoing muons for this data set. We combine
these data with the previously published data (Ahlen, 1995)  for a total of 642 upgoing events.
Based on events outside the upgoing muon peak, we estimate
there are $12.5 \pm 6$ background events in the total data set.
In addition to these events, we estimate
that there are $10.5 \pm 4$ events which result from upgoing
charged particles produced by
downgoing muons in the rock near MACRO.
Finally, it is estimated that $12 \pm 4$ events are the
result of interactions of neutrinos in the very bottom layer of MACRO
scintillators. Hence, removing the backgrounds, the
observed number of upgoing throughgoing muons integrated over all
zenith angles is 607.

In the upgoing muon simulation we have used the neutrino flux 
computed by the Bartol group (Agrawal 1996). 
 The cross-sections for the neutrino interactions have 
been calculated using the GRV94 (Gl\"{u}ck,1995) parton
distributions set, which varies by +1\% respect to the Morfin and Tung 
parton distribution that
 we have used in the past.  We estimate a
systematic error of 9\% on the upgoing muon flux due to
uncertainties in the cross section including low-energy effects (Lipari 1995). The propagation of muons
to the detector has been done using  the energy loss calculation (Lohmann 1985) for standard rock.
The total systematic uncertainty on the expected flux of muons 
 adding the errors from neutrino flux, cross-section 
and muon propagation in quadrature is $\pm17\%$.
This theoretical error in the prediction is mainly a scale error that doesn't change the shape of the angular
distribution.
The number of events expected integrated over all zenith angles is
824.6, giving a ratio of the observed number of events to the expectation 
of 0.74 $\pm0.031$(stat) $\pm0.044$(systematic) $\pm0.12$(theoretical).

Figure 1 B) shows the zenith angle distribution of the measured
flux of upgoing muons with energy greater than 1 GeV for all MACRO data
compared to the Monte Carlo expectation for no oscillations 
 and with a $\nu_{\mu} \rightarrow\nu_{\tau}$ oscillated flux with $\sin^2 2 \theta = 1$ and 
$\Delta m^2 = 0.0025$ eV$^2$ (dashed line).

The shape of the angular distribution has been tested with the hypothesis of no oscillation
excluding the last bin near the horizontal and normalizing data and predictions.
The $\chi^2$ is $22.9$,  for 8 degrees of freedom
(probability of 0.35\% for a shape at least this different from the
expectation). We have considered also oscillations $\nu_{\mu} \rightarrow\nu_{\tau}$.
The best $\chi^2$ in the physical region of the oscillations parameters is 12.5
for $\Delta m^2$ around $0.0025 eV^2$ and maximum mixing (the  best
 $\chi^2$ is 10.6 , outside the physical region for an unphysical 
value of $\sin^2 2 \theta =1.5$).

To test the oscillation hypothesis, we calculate the independent probability for 
obtaining the number of events observed and the angular distribution 
for various oscillation parameters. They are reported for $\sin^2 2 \theta = 1$ in Figure 2 A) for  $\nu_{\mu}
\rightarrow\nu_{\tau}$ oscillations.
 It is notable that the value of
$\Delta m^2$  suggested from the shape of the angular distribution is similar to the value necessary
in order to obtain the observed reduction in the total number of events in the hypothesis 
of maximum mixing. Figure 2 B) shows the same quantities for sterile neutrinos oscillations (Akhmedov
1993,Liu 1998).

Figure 3 A) shows probability contours for oscillation
parameters using the combination of probability for
the number of events and $\chi^2$ of the angular distribution.
The maximum of the probability is 36.6\% for oscillations $\nu_{\mu} \rightarrow\nu_{\tau}$.
The best probability for oscillations into sterile neutrinos is 8.4\%.
The probability for no oscillation is 0.36\%.

Figure 3 B) shows the
confidence regions at the 90\% and 99\% confidence levels based on
application of the Monte Carlo prescription in 
(Feldman 1998).  We plot also  the
sensitivity of the experiment. The sensitivity is the 90\% contour which would result from the preceding
prescription when the data are equal to the Monte Carlo prediction at the
best-fit point.

\begin{figure}
\epsfig{file=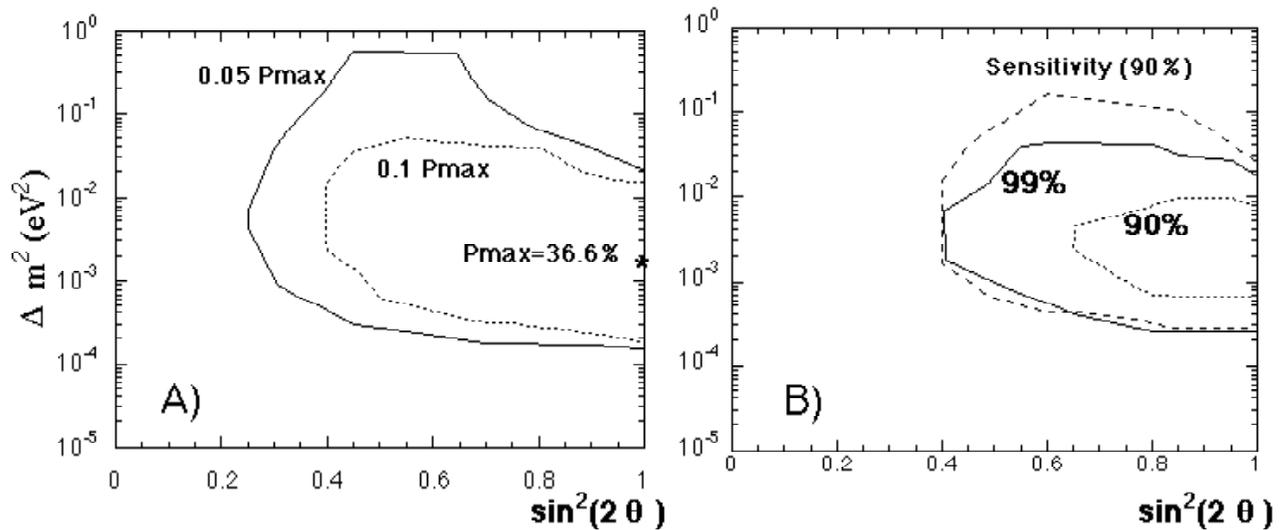,height=70mm}
%\psfig{file=fig4.eps,height=75mm}
%\indent{\picture 110mm by 65mm (Fig4 scaled 620)}
\caption {\label{fig:escl} A) Probability contours
  for oscillation parameters for $\nu_\mu \rightarrow \nu_\tau$
  oscillations based on the combined probabilities of zenith shape and
  number of events tests. 
   The best probability in the physical region is 36.6\% and iso-probability
   contours are shown for 10\% and 5\% of this value (i.e. 3.6\% and
   1.8\%).    B) Confidence regions 
   at the 90\% and 99\% levels calculated according to Feldman 1998.
   Since the best probability is outside the physical region
   the confidence interval regions are smaller than the ones expected
   from the sensitivity of the experiment.}
\end{figure}

\section{Conclusions:}
The upgoing throughgoing muon data set is in favor of $\nu_{\mu} \rightarrow\nu_{\tau}$
oscillation  with  parameters similar to  those observed by Superkamiokande 
with a probability of 36.6\%  against the 0.36\% for the no
oscillation hypothesis. The probability  of oscillations from the angular 
distributions only is 13\%.
The probabilities are higher than the ones of the old data (Ambrosio 1998b).
The neutrino sterile oscillation hypothesis is slightly disfavored. 

%  References: (DO NOT ALTER NEXT 4 LINES)
\vspace{1ex}
\begin{center}
{\Large\bf References}
\end{center}
Agrawal V. et al 1996 Phys. Rev.  D53  1314\\
Ahlen S.  et al.(MACRO collaboration) 1995,  Phys. Lett.  B 357  481\\
Ahlen S. et al.(MACRO collaboration) 1993,  Nucl.Instrum.Meth.A324:337-362 \\
Akhmedov E. ,  Lipari P.  Lusignoli M. 1993, Phys.Lett. B300:128-136 \\
Ambrosio M. et al.(MACRO collaboration) 1998a, Astropart.Phys.9:105-117 \\
Ambrosio M. et al.(MACRO collaboration) 1998b, Phys Lett. B. 434 451 \\
Feldman G.  and Cousins  R. 1998  Phys. Rev.  D57 3873\\
Fukuda Y. et al. (SuperKamiokande collaboration) 1998a Phys.Rev.Lett.81:1562-1567\\
Fukuda Y. et al. (SuperKamiokande collaboration)  1998b, e-Print Archive hep-ex/9812014 \\
Gl\"{u}ck M., Reya E. and Stratmann M.1995, Z. Phys. C67, 433 \\
Hatakeyama S.  et al. (Kamiokande collaboration) 1998, Phys Rev Lett 81 2016 \\
Lipari P.  Lusignoli M. and  Sartogo F. 1995, Phys. Rev. Lett.  74  4384 \\
Liu Q.Y. and   Smirnov A.Yu. 1998, Nucl.Phys. B524  505 \\
Lohmann H. Kopp R.,Voss R.  1985, CERN-EP/85-03\\
Surdo A. (MACRO collaboration) 1999, HE4.1.06 in this conference
\end{document}